\documentclass[iop]{emulateapj}

\usepackage{graphicx}
\usepackage[pdf]{pstricks}
\usepackage{color}

\slugcomment{Accepted for Publication in Astrophysical Journal Letters}

\shorttitle{M87 Jet Optical Proper Motions}
\shortauthors{Meyer et al.}

\begin{document}

\title{Optical Proper Motion Measurements of the M87 Jet: New Results from the Hubble Space Telescope}

\author{Eileen T. Meyer\altaffilmark{1}}
\author{W. B. Sparks\altaffilmark{1}}
\author{J. A. Biretta\altaffilmark{1}}
\author{Jay Anderson\altaffilmark{1}}
\author{Sangmo Tony Sohn\altaffilmark{1}}
\author{Roeland P. van der Marel\altaffilmark{1}}
\author{Colin Norman\altaffilmark{1,2}}
\author{Masanori Nakamura\altaffilmark{3}}

\email{Email: meyer@stsci.edu}
\altaffiltext{1}{Space Telescope Science Institute, 3700 San Martin Drive, Baltimore, MD 21218, USA}
\altaffiltext{2}{Department of Physics and Astronomy, Johns Hopkins University, Baltimore, MD 21218, USA}
\altaffiltext{3}{Institute of Astronomy and Astrophysics, Academia Sinica, P.O. Box 23-141, Taipei 10617, Taiwan}

\begin{abstract}
We report new results from an HST archival program to study proper
motions in the optical jet of the nearby radio galaxy M87. Using over
13 years of archival imaging, we reach accuracies below
0.1$c$ in measuring the apparent velocities of individual knots in the
jet. We confirm previous findings of speeds up to 4.5$c$ in the inner
6\arcsec\, of the jet, and report new speeds for optical components in
the outer part of the jet. We find evidence of significant motion
transverse to the jet axis on the order of 0.6$c$ in the inner jet
features, and superluminal velocities parallel and transverse to the
jet in the outer knot components, with an apparent ordering of
velocity vectors possibly consistent with a helical jet
pattern. Previous results suggested a global deceleration over the
length of the jet in the form of decreasing maximum speeds of knot
components from HST-1 outward, but our results suggest that
superluminal speeds persist out to knot C, with large differentials in
very nearby features all along the jet. We find significant
apparent accelerations in directions parallel and transverse to the jet axis,
along with evidence for stationary features in knots D, E, and
I. These results are expected to place important constraints on
detailed models of kpc-scale relativistic jets.
\end{abstract}

\keywords{galaxies: jets --- galaxies: individual(M87) --- galaxies: active --- astrometry}

\section{Introduction}
The nearby radio galaxy M87, a giant elliptical near the center of the
Virgo cluster, hosts a striking optical jet extending 20\arcsec\, to
the northwest of a blazar-like core, as first observed by
\cite*{cur18}. This source 
has been extensively observed in the radio, infrared, optical, and
X-rays
\citep[e.g.,][]{owen89_rad,sparks96_opt,per01_IR,walker08_vlbi,marshall2002_xray,perlman2005_xray}. At
a distance of only 16.7 Mpc \citep[81 pc/\arcsec,][]{bla09},
structures on the order of parsecs can easily be resolved using
high-resolution instruments such as the Hubble Space Telescope
(HST). The bright knots of emission spread along the 1.6 kpc
(projected) jet in optical and X-ray imaging correspond to features
seen in the radio, with a spectrum consistent with synchrotron
emission \citep{marshall2002_xray}.


Proper motion studies are integral to our understanding of
relativistic jets, as apparent velocities ($\beta_\textrm{app}$=v/$c$)
give lower limits to the bulk Lorentz factor and upper limits on the
angle to the line of sight, as well as detailed kinematics for
comparison to theoretical jet models. Almost all proper motion studies
of jetted AGN have used Very Long Baseline Interferometry (VLBI),
and in the case of M87 have revealed sub-relativistic speeds in small
components within a parsec of the core
\citep{reid89,junor95,kovalev07}. \emph{Superluminal} motions have
also been measured in the kpc-scale jet with both the VLA
\citep[][hereafter B95]{bir95}, and in the optical with HST
\citep[][hereafter B99]{bir99}. The latter study measured only the
inner 6\arcsec of the jet, and suffered from rather larger errors (on
the order of 0.3 - 1$c$). However, in comparison to radio
measurements, the optical jet emission traces higher energy electron
populations more closely associated with the sites of acceleration,
leading to more compact features which can be precisely measured.

In this letter we present the initial results of an archival study of
proper motions in the M87 jet with HST. The aim of this work is to
leverage 13.25 years of Hubble observations of M87 to vastly improve
on previous measurements of apparent velocities in the jet, reaching
accuracies better than 0.1$c$ in order to map the detailed velocity field of
the jet as it propagates from the core, including transverse motions
and accelerations.


\begin{figure*}
\centering
\includegraphics[width=6.5in]{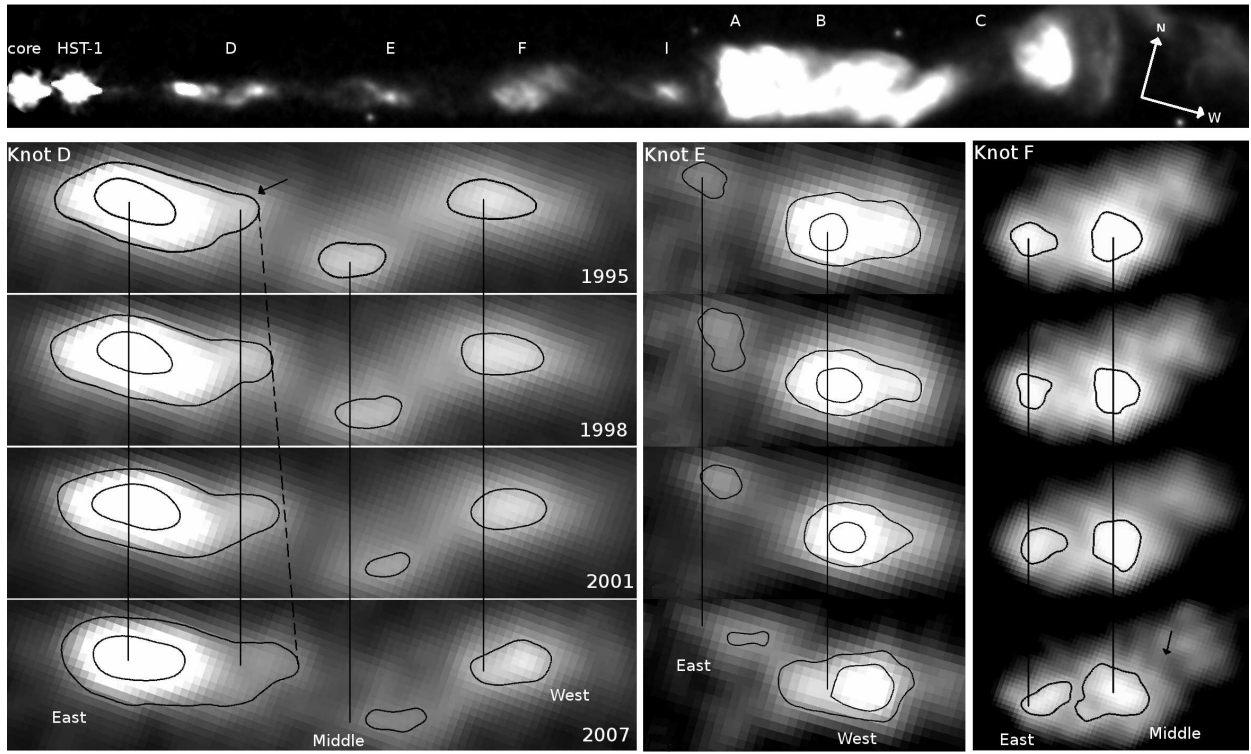}
\caption{\emph{(Top Figure)} The entire jet, from a stack of ACS/WFC
  F606W images taken in 2006 with galaxy subtracted. All other images
  are taken from stacks of WFPC2 exposures in the F814W
  filter. \emph{(Lower Left)} Within knot D, D-East appears to be
  nearly stationary, while D-Middle appears to move at
  4.27$\pm$0.30$c$ along the jet. D-West is the only knot to show a
  strong deceleration in the jet-direction, apparent when comparing
  the 2001 and 2007 epochs. The new component noted with an arrow
  appears to move at $\sim$2.4$c$ (Vertical lines in all figures are
  to guide the eye). \emph{(Lower Middle)} Knot E appears complex,
  with a new extension to the East of the main knot (E-West) appearing
  in 2008. 
  \emph{(Lower Right)} Knot F shows evidence of complex, even clumpy
  structure. F-Middle is moving at 0.36$\pm$0.14$c$ and appears
  to be fading over time. }
\label{fig:innerjet}
\end{figure*}

\section{Observations}
\subsection{Hubble Archival Data}

Because jet components are known to differ in size, intensity, and
apparent location with wavelength, we used archival images
only in a single filter, F814W (wide I band), which gives the longest
possible baseline of 13.25 years and a dense sampling in time. The
total dataset is comprised of nearly all the ACS/HRC, ACS/WFC, and
WFPC2 (PC chip) archival images in this filter. For ACS/HRC, we used
30 (48-50s) exposures spanning November 2002 to November 2006. For
ACS/WFC, we used 217 dithered exposures (all 360s) taken in 2006. For
WFPC2, we used 161 exposures of 160-600s, spanning 1995 to 2008;
several shorter exposures of 30-40s were not used due to poor
signal-to-noise ratio.

\subsection{Astrometry Methods}

The general method we have used is similar to that used (and described
in detail) in \cite*{and10} and \cite*{sohn12}. The reference
frame is based on the positions of hundreds of bright globular
clusters associated with the host galaxy which are easily detected in
the ACS/WFC images and are effectively stationary to proper motions
over the time of this study. As cluster positions do not change with
filter, the reference frame was built using 56 (500s) ACS/WFC
exposures of the field in the F606W filter, which has slightly better
resolution than F814W. The reference frame was created by first detecting the
positions of the globular clusters in each flat-fielded, CTE-corrected
ACS/WFC image\footnote{ACS/HRC and WFPC2 data were not used due to
  inferior in signal-to-noise quality to the longer ACS/WFC images,
  and to avoid complications of combining different instruments.}
using a PSF-peak-fitting routine, then applying the standard
(filter-specific) geometric correction to those positions, and then
finding the best linear transformation for each image which matches
the positions in the individual (geometrically-corrected) frames to a
master reference frame. This last step is done by a routine similar to
MultiDrizzle, but better optimized for astrometry. The process of
finding transformations for all frames is iterated so that the master
reference frame (super-sampled to a pixel scale of 0.025\arcsec/pixel)
effectively gives the average position of each reference source in a
geometrically-corrected frame.

The final reference system consists of positions and magnitudes for
1378 globular clusters within $\sim$100\arcsec of the M87 core
position. Among all clusters, the median one-dimensional RMS residual
relative to the mean position was 0.05 (reference-frame) pixels (1.25
mas), corresponding to a systematic astrometric accuracy ($\times$
1/$\sqrt 56$) of 0.17 mas; over the 13.25 year baseline this is
equivalent to 0.003$c$.

Astrometric solutions were then found for all the ACS/HRC , ACS/WFC,
and WFPC2 images in F814W, except that the linear transformation
solution between the individual (corrected) images and the reference
frame was found once rather than iterated. Typical numbers of globular
clusters used to match the frames were $\sim$200-500 for ACS/WFC,
$\sim$15-30 for WFPC2 images, and $\sim$15-25 for ACS/HRC;
corresponding systematic errors were $\sim$0.006, $\sim$0.03, and
$\sim$0.05 pixels, or 0.003$c$, 0.015$c$, 0.025$c$ over 13.25 years,
respectively.

\subsection{Measuring the Jet Knot Positions} 
Because of the difficulty in measuring the diffuse/complex knot
structures particularly in the noisy HRC images and shorter WFPC2
images, we chose to first identify the peaks of interest using stacked
images. The image stacks include the 2006 ACS/WFC stack, four WFPC2
stacks (epochs 1995-1996, 1998-1999, 2001, and 2007-2008), and three
stacks of ACS/HRC exposures (November 2002 - November 2003, July 2004
- September 2005, November 2005 - November 2006). The host galaxy was
modeled using the ACS/WFC stacked image with the IRAF/STSDAS tasks
\emph{ellipse} and \emph{bmodel} and then subtracted. Seven jet
regions were cut out from these images, corresponding to knots
HST-1, D, E, F, I, A+B, and C (see Figure~\ref{fig:innerjet}).

A two-dimensional continuous functional representation of the image
cutouts was then created using the Cosine Transform function
(\emph{FourierDCT}) in \emph{Mathematica}, which allowed us to find
prominent peaks, as well as contours of constant flux level around
those peaks.  In general, the number of interesting peaks was chosen
by hand, and the contour line levels were at 50\% of the flux of the
peak, after a `background' level was subtracted, the latter estimated
as the minimum flux level between the particular peak and the next
nearest peak. The positions defining the contour were
`reverse-transformed' from master frame coordinates into each
distorted, galaxy-subtracted image so that an intensity-weighted
centroid position from the pixels within the contour could be
calculated.  These final positions were then transformed back into the
common reference frame, so that each peak of interest was measured in
every exposure, resulting in hundreds of position measurements spread
over the 13 year baseline.

\begin{figure}
\centering
\includegraphics[width=3.5in]{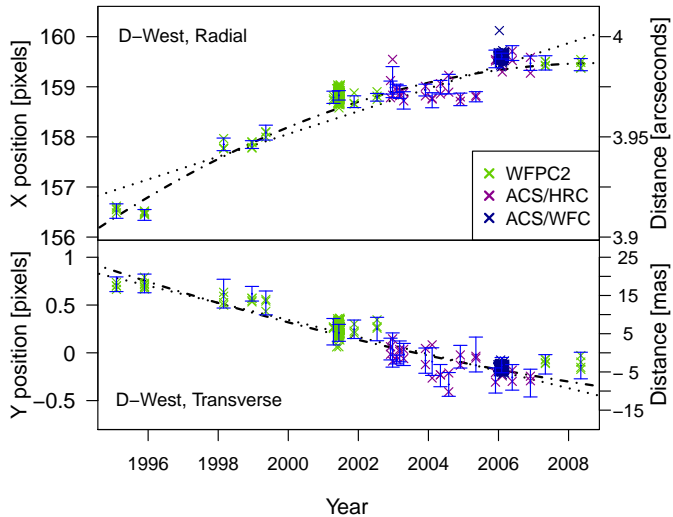}
\caption{\emph{(Upper Panel)}: Distance from the core versus time for
  knot D-West. The blue points show the mean values for each time bin
  with errors estimated from nearby globular clusters. This knot shows
  a remarkable deceleration over time, from an initial speed of
  $\sim$2.7$c$ in 1996 to near zero by the end of the time
  series.\emph{(Lower Panel)} The rapid transverse velocity of knot
  D-West is apparent in this plot of y-coordinate versus time,
  corresponding to $-$0.59$c$.}
\label{fig:dwest}
\end{figure}

To measure the apparent motion of the jet components over time, we
define the x-direction as the line from the core through the center of
knot I (PA of 290$^\circ$), with y-direction perpendicular. Data were binned into 26 time
bins (depicted in Figure~\ref{fig:dwest}), and a linear model was fit using standard
weighted least squares.  The weights were taken to be the inverse of
the variance for each time bin, as measured from the 10 nearby
globular cluster reference sources (discussed below). For each feature
and in each direction, we also attempted to fit the data with a
quadratic to look for accelerations.  When the Analysis of Variance
(ANOVA) favored the quadratic over the linear model at a 95\% level or
greater, the quadratic fit is noted in Table~1 along with the linear
fit.  The apparent speed ($\beta_\mathrm{app}$) in each
direction in units of $c$ is also given; where the quadratic fit is
favored $\beta_\mathrm{app}$ corresponds to the speed at the mean date
of 2004.06.

\subsection{The Globular Clusters as Controls}

The method was tested using ten globular clusters near to the jet,
each of which was measured as described above, with contour levels
drawn at 50\% of the peak intensity. Each set of cluster positions was
binned into the 26 time bins spread between 1995 and 2008, and for
each bin the mean position for all ten clusters was subtracted from
the positions so that the variance in the estimates could be measured
from all clusters as a function of epoch for use in the weighted
linear least squares fit. The measured variance was essentially
constant across all epochs at $\sim$0.1 reference pixels. Linear fits
to the x- and y-direction positions versus time in all cases were
consistent with a slope of zero, with typical errors on the order of
0.003 pixels yr$^{-1}$ or 0.02$c$. 

\section{Results}
\begin{figure*}
\centering
\includegraphics[width=6.7in]{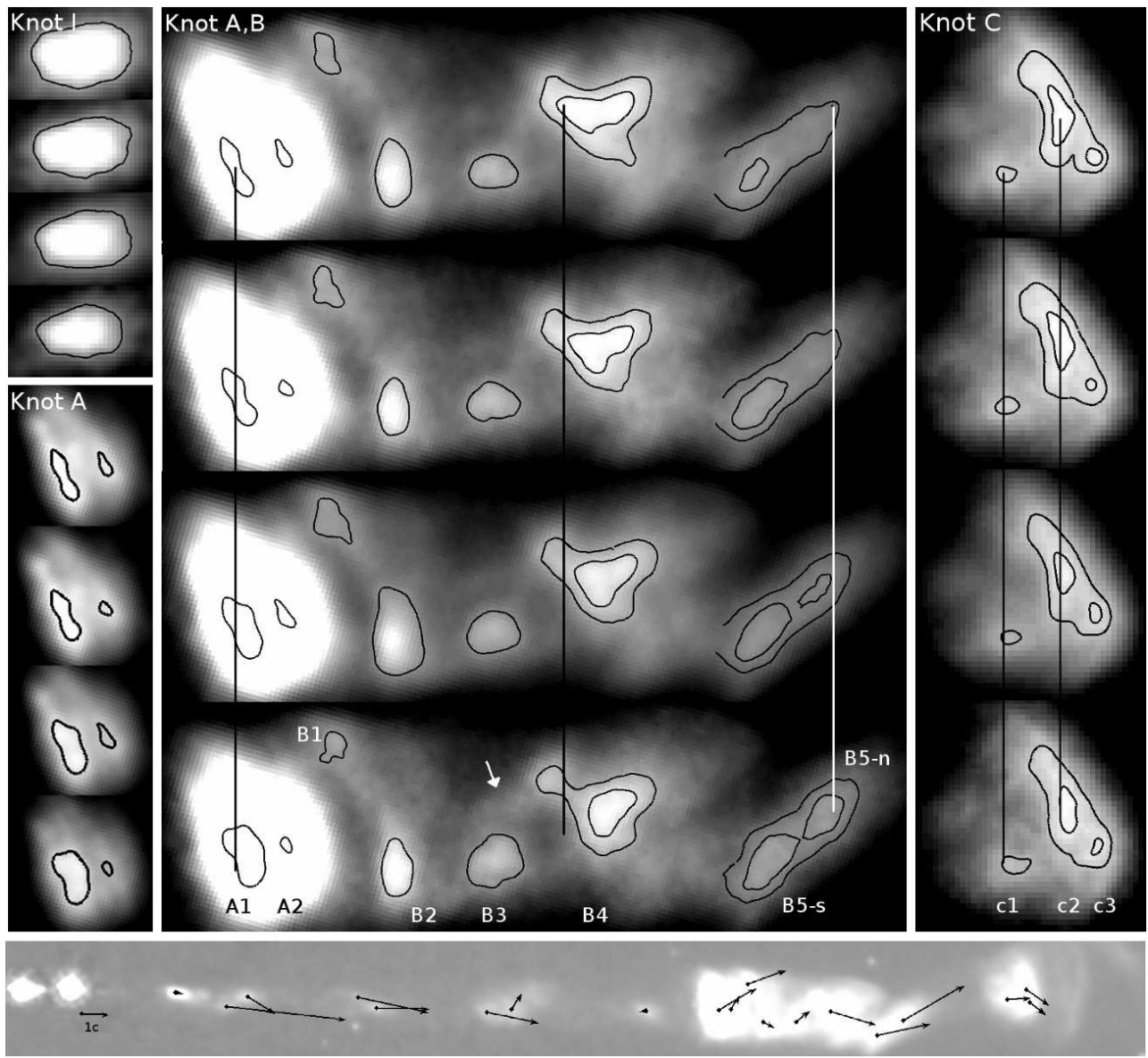}
\caption{\emph{(Upper Left)} Knot I appears to fade and move backwards
  along the jet at 0.23$\pm$0.12$c$. \emph{(Lower Left}) Knot A is
  shown with a stretch to emphasize the fading of the knot over
  time. \emph{(Middle Panel)} The knot A/B complex shows remarkable
  variability, with both sub-relativistic and superluminal apparent
  motions. The white arrow is pointing to a `bar' which appears in the
  last epoch. \emph{(Right Panel)} Knot C shows speeds on the order of 0.5-1$c$. \emph{(Bottom Panel)} A depiction
  of velocities as vectors from their positions along the jet.}
\label{fig:outerjet}
\end{figure*}

Starting from the core, the first well-detected features in our
dataset include a small narrow extension from the core (out to
$\sim$0.5\arcsec) probably corresponding to knot L seen in VLBI
images, as well as the famous variable feature, HST-1. Both the core
and HST-1 are highly variable in flux and frequently saturated in our
images. Because special care must be taken with saturated images, an
analysis of the inner region of the M87 jet up to and including HST-1
will be addressed in a follow-up to this letter. For all the knots
described below, the WFPC2 image stacks (galaxy
subtracted) described above are shown in Figures~\ref{fig:innerjet}
and \ref{fig:outerjet}.\footnote{Movies of the jet and knots are also
  available in the online version of this paper.}.

\subsection{The Inner Jet: Knots D, E, F, and I}

Knot D is the most consistently bright feature after the knot A/B/C
Complex, and significant proper motions were measured in both previous
studies.\footnote{It is important to note that B95 used data from 1985
  through 1995, and so directly precedes the epoch of our dataset,
  while B99 overlaps with only the early part of it, spanning
  1995-1998. All $\beta_\mathrm{app}$ (including previous work) were
  computed with the conversion factor 0.264$c$ yr mas$^{-1}$.} As
suggested by the contours overlaid on Figure~\ref{fig:innerjet},
D-Middle is one of the fastest components with a speed of
4.27$\pm$0.30$c$ along the jet, while the brighter D-West shows
evidence of \emph{deceleration} radially, slowing to a near stop by
the final epoch in 2008, while maintaining one of the largest
transverse speeds of $-$0.59$\pm$0.05$c$ (see Figure~\ref{fig:dwest}).
These measurements are consistent with the results of B99
(5.26$\pm$0.92$c$ and 2.77$\pm$0.98$c$ for D-Middle and D-West,
respectively), where we compare with our estimated velocity of D-West
in 1996 of 2.73$\pm$0.40$c$ from the quadratic fit.

Previous results on knot D-East have been conflicting: B95 found that
it moved \emph{inward} along the jet at 0.23$\pm$0.12$c$ (possibly
consistent with being stationary), while B99 found a large outward
apparent velocity of 3.12$\pm$0.29$c$; our result of 0.28$\pm$0.05 is
more in line with B95. It is possible that there is a stationary
feature at D-East, through which components emerge (analogous to what
is seen in HST-1). In that case, the higher-resolution FOC was perhaps
tracking an emerging bright component, while over longer periods the
global feature at D-East is stationary.

As shown in the central lower panel of Figure~\ref{fig:innerjet}, the
main knot structure of knot E (E-West) is complex, with a moderately
superluminal speed of 1.91$\pm$0.14$c$, about half of that found in
B99 (4.07$\pm$0.85$c$).  In addition to the 50\% intensity contrast
contour line shown, a lower-intensity contour at 30\% has been drawn
around this area, to illustrate the morphological changes of this
feature over time. To the west there appears a bright extension in the
1995 epoch which advances westward faster than the brightest part, and
rapidly fades in the later epochs. Interestingly, a new component
appears to the East in 2008, suggesting again the possibility of a
favored location or stationary feature, which should be checked with
future observations.


Like knot E, the next feature in the jet, knot F, is diffuse, faint,
and apparently complex. We track two broad features, F-East and
F-Middle, as shown in the lower right of Figure 1. As B99 were unable
to measure proper motions for anything beyond knot E, the only
previous measurement gives 0.90$\pm$0.23$c$ in B95, for the entire
diffuse feature seen in the radio. This is slightly faster than the
value we measure for the most prominent component F-Middle
(0.36$\pm$0.14$c$), however, the epochs of observation are not
contemporaneous and slight deceleration is possible.


Finally, knot I has a slight negative velocity along the jet axis, of
$-$0.23$\pm$0.12$c$. The knot clearly fades steadily with time, as
shown in the upper left of Figure 3. This knot is the clearest case of
a possible stationary feature in the jet outside of the upstream component
HST-1d which has been extensively observed with VLBI
\citep[$\beta_\mathrm{app}<$0.25$c$;][]{cheung07}. 

\subsection{The Outer Jet: Knots A, B, and C}

The outer part of the jet looks very different from the evenly spaced
knots discussed previously.  Knot A is the brightest feature in the
entire jet, excepting HST-1 during its remarkable flare of 2002 - 2008
\citep{har06}. It has been suggested that knot A is an oblique shock
which precipitates the break-up of the orderly structure of the jet,
thus resulting in the complex knot B region downstream, and the final
knot C feature which is considerably off the main jet axis
\citep{bic96}. Within the bright but extended knot A, we track two
features which appear on top of this more diffuse emission (see
Figure~\ref{fig:outerjet}). For A1, we find a relativistic velocity of
1.32$\pm$0.12$c$, however it is not clear that this feature is really
a single component, as the western side of it appears to move slowly
if at all. For A2, we measure a much slower radial velocity of
0.31$\pm$0.06$c$, consistent with the previous results of 0.4-0.6$c$
from B95.

Qualitatively, the A/B complex is surprisingly `active' with new
features brightening up and moving with high speeds relative to other
nearby components. A good example is pointed out by the arrow in the
last epoch of the central panel in Figure~\ref{fig:outerjet}, where a
`bar' has formed between B3 and B4. At the same time, within region B4
a separation is taking place, leaving a small knot behind to the
northeast. While B4 is probably a complex of at least 2-3 `subknots',
we used the inner contour (at 80\% median flux), giving us an apparent
speed of 1.66$\pm$0.11$c$ along the jet direction. B1,B2, and B3 as
well as the final knot C show significant apparent velocities in both
directions, on the order of 0.5-1.5$c$.

The final terminus of knot B is particularly interesting, as a new
distinct component (which we have labeled B5-n) appears in the
northern half in the 2001 epoch. Using data from 2001 - 2008 only (the
time when this component was detected), we measure an apparent
velocity of 2.16$\pm$0.35$c$ along the jet direction and
1.30$\pm$0.31$c$ in the transverse direction.  Qualitatively, the
northern end of B5 seems to continuously brighten while the whole
feature stretches up towards the northwest. In knot B generally there
is a repeating motif of two bright regions connected by a fainter
`bar'; this is apparent between knots B1 and B2, between B3 and the
remnant of B4, and the two halves of B5.

\begin{figure}
\centering
\includegraphics[width=3.2in]{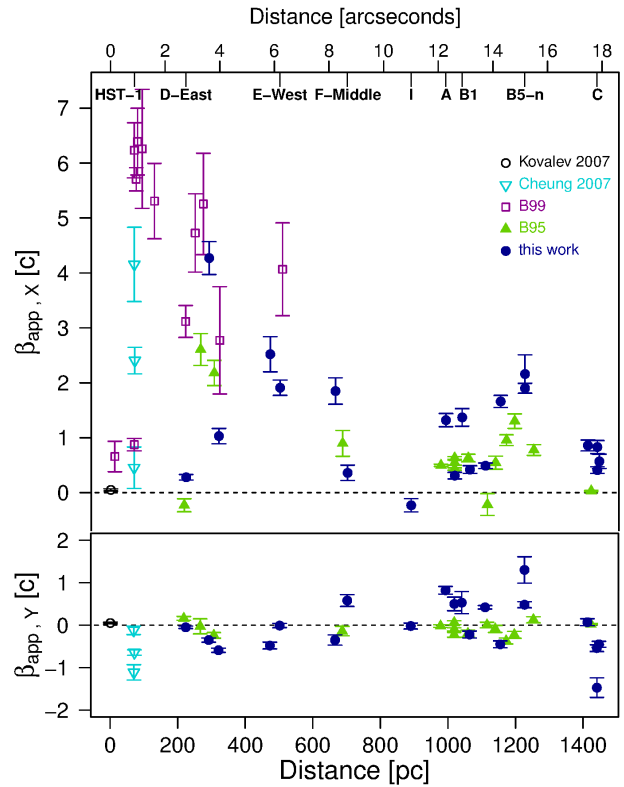}
\caption{Velocities along the jet (upper panel) and transverse to the
  jet (lower panel) as a function of distance from the core. Previous
  measurements shown for comparison taken from B95, B99,
  \citealt{cheung07}, and \citealt{kovalev07}.}
\label{fig:envelope}
\end{figure}

The most striking finding of the outer jet component velocities is
depicted in the lower panel of Figure~\ref{fig:outerjet}, where the
velocity vectors are plotted from the center of each knot.  There is a
conspicuous `tip-to-tail' alignment of almost all the vectors within
the outer knot A/B/C complex, strongly suggestive of a flattened view
of a helical motion which might result in such a `zig-zag' pattern,
though this should be checked with theoretical modeling.

\section{Discussion and Conclusions}
\label{sec:final}

We have presented new proper motion measurements for the knots in the
M87 optical jet, reaching accuracies better than 0.1$c$. The speeds
are largely in agreement with previous results where they exist (see
Figure~\ref{fig:envelope}); some discrepancies are likely due to the
higher resolution in the optical allowing us to track more compact
features than those measured in the radio, or changes such as the
observed deceleration of knot D-West. We also find evidence of
stationary features in knots D-East, I, and possibly E (if the new
extensions in 2008 is a recurrence at a preferred location), which
should be checked with future observations. Both D-Middle, and the
small component extending from D-East (labeled with an arrow in
Figure~\ref{fig:innerjet}) appear to steadily drop in intensity over
the 13 years of observation, a very similar behavior to the western
components of knot HST-1 (e.g., B99), and both HST-1 and knot D are
dominated by a transverse magnetic field as noted by
\citep{perlman99_pol}, which suggests a shock at these locations. It
has been suggested that at the eastern edge of HST-1 there is a
recollimation shock (\citealt{sta06,cheung07}; see also
\citealt{asada2012}), but the appearance of very different speeds all
along the jet might also be consistent with pairs of forward/reverse
fast-mode MHD shocks in a strongly magnetized relativistic flow
\citep{nakamura2010}.

In the outer jet (from knot A), we find apparent velocities that are still
superluminal and velocity vectors that appear to line up into a
helical/side-to-side pattern. However it is not clear that the
measured speeds can be identified with discrete moving components. It
is possible that if the jet of M87 has a helical structure which is
broken up at knot A, we are seeing the `unwinding' of the coil in knot
B. Thus the appearance of the bar noted with an arrow in 2008 could be
the result of a change in geometry as the coil of the jet unwinds,
bringing a length into our line of sight. Sideways oscillation of
components B1-B5 may presumably follow a helical path of underlying
magnetic fluxes \citep{owen89_rad,perlman99_pol} in three
dimensions. Thus, the knot A/B/C complex could be seen as a
consequence of the interplay between the slow-mode MHD shocks and
growing helical kink instability \citep{nakamura_2004_hk}.

\acknowledgments This project is part of the HSTPROMO collaboration
(www.stsci.edu/~marel/hstpromo.html), a set of HST projects aimed at
improving our dynamical understanding of stars, clusters, and galaxies
in the nearby Universe through measurement and interpretation of
proper motions. Support for this work was provided by NASA through a
grant for archival program 12635 from the Space Telescope Science
Institute (STScI), which is operated by the Association of
Universities for Research in Astronomy (AURA), Inc., under NASA
contract NAS5-26555. {\it Facilities:} \facility{HST (WFPC2, ACS)}.

\clearpage

\begin{deluxetable}{clcccccccccccc}

  \tabletypesize{\footnotesize}
\tablecolumns{14} 
\tablewidth{0pc} 
\tablehead{ 
\colhead{}        &  \colhead{}         & \colhead{} & \colhead{}                                  &   \multicolumn{2}{c}{Linear Fit}        & \colhead{} & \multicolumn{4}{c}{Quadratic Fit}                                            & \colhead{} & \multicolumn{2}{c}{Speed} \\
                                                                                                       \cline{5-6}                                          \cline{8-11}                                                                   \cline{13-14} \\ 
\colhead{Knot}    & \colhead{Subknot}   & \multicolumn{2}{c}{Distance\tablenotemark{$\dagger$}}    &   \colhead{$\mu_x$} & \colhead{$\mu_y$}  & \colhead{} & \colhead{$\mu_x$} & \colhead{$\xi_x$} & \colhead{$\mu_y$} & \colhead{$\xi_y$} & \colhead{} & \colhead{$\beta_x$} &  \colhead{$\beta_y$} \\
& & \arcsec & kpc & mas yr$^{-1}$ & mas yr$^{-1}$ & \colhead{} & mas yr$^{-1}$ & mas yr$^{-2}$ & mas yr$^{-1}$ & mas yr$^{-2}$ & \colhead{} & $c$ & $c$}
\startdata 
D  & East     & \phantom{1}2.77 & 0.224 &    \phantom{$-$0}1.1$\pm$0.2  &            $-$0.2$\pm$0.1\phantom{0}   & &                           &                    &                              &                                 & &   \phantom{$-$}0.28$\pm$0.05 &            $-$0.05$\pm$0.03 \\
   & Middle\tablenotemark{a}   & \phantom{1}3.61 & 0.292 &    \phantom{$-$}16.2$\pm$1.1  &            $-$0.7$\pm$0.1\phantom{0}   & &                           &                    &              $-$1.3$\pm$0.2  &             $-$0.2$\pm$0.1      & &   \phantom{$-$}4.27$\pm$0.30 &            $-$0.35$\pm$0.05 \\
   & West     & \phantom{1}3.97 & 0.321 &    \phantom{$-$0}5.7$\pm$0.5  &            $-$2.2$\pm$0.2\phantom{0}   & & \phantom{$-$}3.9$\pm$0.5  & $-$0.80$\pm$0.2    &                              &                                 & &   \phantom{$-$}1.03$\pm$0.14 &            $-$0.59$\pm$0.05 \\[2ex]

E  & East     & \phantom{1}5.86 & 0.473 &    \phantom{$-$0}9.6$\pm$1.2  &            $-$1.8$\pm$0.3\phantom{0}   & &                           &                    &                              &                                 & &   \phantom{$-$}2.52$\pm$0.32 &            $-$0.48$\pm$0.08 \\
   & West     & \phantom{1}6.21 & 0.502 &    \phantom{$-$0}7.2$\pm$0.5  &  \phantom{$-$}0.0$\pm$0.2\phantom{0}   & &                           &                    &                              &                                 & &   \phantom{$-$}1.91$\pm$0.14 &            $-$0.01$\pm$0.05 \\[2ex]

F  & East     & \phantom{1}8.23 & 0.666 &    \phantom{$-$0}7.0$\pm$0.9  &            $-$1.3$\pm$0.4\phantom{0}   & &                           &                    &                              &                                 & &   \phantom{$-$}1.85$\pm$0.24 &            $-$0.35$\pm$0.12 \\
   & Middle   & \phantom{1}8.68 & 0.702 &    \phantom{$-$0}1.4$\pm$0.5  &  \phantom{$-$}0.7$\pm$0.5\phantom{0}   & &                           &                    &    \phantom{$-$}2.2$\pm$0.5  &   \phantom{$-$}0.6$\pm$0.2      & &   \phantom{$-$}0.36$\pm$0.14 &  \phantom{$-$}0.58$\pm$0.14 \\[2ex]

I  &          &           11.01 & 0.890 &    \phantom{0}$-$0.9$\pm$0.4  &            $-$0.1$\pm$0.2\phantom{0}   & &                           &                    &                              &                                 & &             $-$0.23$\pm$0.12 &            $-$0.02$\pm$0.07 \\[2ex]

A  & A1       &           12.28 & 0.993 &    \phantom{$-$0}5.0$\pm$0.4  &  \phantom{$-$}2.4$\pm$0.3\phantom{0}   & &                           &                    &    \phantom{$-$}3.1$\pm$0.3  &   \phantom{$-$}0.3$\pm$0.1      & &   \phantom{$-$}1.32$\pm$0.12 &  \phantom{$-$}0.82$\pm$0.09 \\
   & A2       &           12.60 & 1.019 &    \phantom{$-$0}1.2$\pm$0.2  &  \phantom{$-$}1.9$\pm$0.6\phantom{0}   & &                           &                    &                              &                                 & &   \phantom{$-$}0.31$\pm$0.06 &  \phantom{$-$}0.50$\pm$0.16 \\[2ex]

B  & B1       &      12.88      & 1.041 &    \phantom{$-$0}5.2$\pm$0.6  &            $-$0.6$\pm$0.9\phantom{0}   & &                           &                             &    \phantom{$-$}2.0$\pm$1.0  &   \phantom{$-$}1.1$\pm$0.3      & &   \phantom{$-$}1.37$\pm$0.16 &  \phantom{$-$}0.53$\pm$0.26 \\
   & B2       &      13.16      & 1.064 &    \phantom{$-$0}1.6$\pm$0.3  &            $-$1.4$\pm$0.2\phantom{0}   & &                           &                             &              $-$0.8$\pm$0.3  &   \phantom{$-$}0.2$\pm$0.1      & &   \phantom{$-$}0.42$\pm$0.07 &            $-$0.22$\pm$0.08 \\
   & B3       &      13.74      & 1.110 &    \phantom{$-$0}1.3$\pm$0.2  &  \phantom{$-$}1.2$\pm$0.1\phantom{0}   & &  \phantom{$-$}1.9$\pm$0.2 & \phantom{$-$}0.3$\pm$0.1    &    \phantom{$-$}1.6$\pm$0.2  &   \phantom{$-$}0.2$\pm$0.1      & &   \phantom{$-$}0.49$\pm$0.05 &  \phantom{$-$}0.42$\pm$0.04 \\
   & B4       &      14.29      & 1.155 &    \phantom{$-$0}6.3$\pm$0.4  &            $-$3.3$\pm$0.5\phantom{0}   & &                           &                             &              $-$1.7$\pm$0.3  &   \phantom{$-$}0.7$\pm$0.1      & &   \phantom{$-$}1.66$\pm$0.11 &            $-$0.45$\pm$0.08 \\
   & B5-s     &      15.18      & 1.227 &    \phantom{$-$0}7.2$\pm$0.3  &  \phantom{$-$}2.8$\pm$0.3\phantom{0}   & &                           &                             &    \phantom{$-$}1.8$\pm$0.3  &             $-$0.4$\pm$0.1      & &   \phantom{$-$}1.90$\pm$0.09 &  \phantom{$-$}0.48$\pm$0.07 \\
   & B5-n     &      15.18      & 1.227 &    \phantom{$-$0}8.2$\pm$1.3  &  \phantom{$-$}4.9$\pm$1.2\phantom{0}   & &                           &                             &                              &                                 & &   \phantom{$-$}2.16$\pm$0.35 &  \phantom{$-$}1.30$\pm$0.31 \\[2ex]

C  & C1       &      17.48      & 1.413 &    \phantom{$-$0}3.3$\pm$0.4  &            $-$0.6$\pm$0.3\phantom{0}   & &                           &                             &    \phantom{$-$}0.3$\pm$0.3  &   \phantom{$-$}0.4$\pm$0.1      & &   \phantom{$-$}0.86$\pm$0.10 &  \phantom{$-$}0.07$\pm$0.08 \\
   & C2       &      17.83      & 1.441 &    \phantom{$-$0}2.4$\pm$0.3  &            $-$1.7$\pm$1.2\phantom{0}   & &  \phantom{$-$}1.5$\pm$0.2 &        $-$0.4$\pm$0.1       &              $-$5.6$\pm$0.9  &             $-$1.6$\pm$0.3      & &   \phantom{$-$}0.41$\pm$0.06 &            $-$1.47$\pm$0.23 \\
   & C3       &      17.90      & 1.447 &    \phantom{$-$0}2.2$\pm$0.5  &  \phantom{$-$}0.7$\pm$0.6\phantom{0}   & &                           &                             &              $-$1.7$\pm$0.3  &             $-$1.0$\pm$0.1      & &   \phantom{$-$}0.57$\pm$0.13 &            $-$0.45$\pm$0.07 \\
   & C2+C3    &      17.83      & 1.441 &    \phantom{$-$0}3.2$\pm$0.4  &            $-$2.1$\pm$0.3\phantom{0}   & &                           &                             &                              &                                 & &   \phantom{$-$}0.83$\pm$0.12 &            $-$0.54$\pm$0.08 \\[2ex]

GC\tablenotemark{b} &   (single)&     & &       0.0$\pm$0.22           &             0.31$\pm$0.15\phantom{0}   & &                           &                             &                              &                                 & &   \phantom{$-$}0.00$\pm$0.06 &            0.08$\pm$0.04 \\
                    &   (mean)  &     & &       0.0$\pm$0.06           &             0.00$\pm$0.06\phantom{0}   & &                           &                             &                              &                                 & &   \phantom{$-$}0.00$\pm$0.02 &            0.00$\pm$0.02 \\
\enddata 
\tablecomments{Quadratic fit given when preferred at $>$95\% level, where $\beta$ values in those cases correspond to the mean time 2004.06. No HRC data was used for knots D-Middle, F,A,B, and C due to poor signal-to-noise.}
\tablenotetext{$\dagger$}{Distance from the core as measured in the 1995 epoch WFPC2 stacked image.}
\tablenotetext{a}{No data after 2007 used in fit due to feature fading.}
\tablenotetext{b}{Examples of Globular Cluster (GC) fits, including a single faint GC near knot A and the ensemble average.}
\end{deluxetable}

\end{document}